\documentclass{llncs}
\usepackage{makeidx}
\usepackage[justification=centering]{caption}
\usepackage[noend]{algpseudocode}
\usepackage{algorithm}
\usepackage{amsmath}
\usepackage{amssymb}
\usepackage{blindtext}
\usepackage{booktabs}
\usepackage{caption}
\usepackage{changes}
\usepackage{cite}
\usepackage{esvect}
\usepackage{graphicx}
\usepackage{hyperref}
\usepackage{pgfplots}
\usepackage{pifont}
\usepackage{subcaption}
\usepackage{todonotes}
\usepackage{xcolor}

\frontmatter
\begin{document}
\title{
	Ball*-tree: Efficient spatial indexing for constrained nearest-neighbor search in metric spaces
}

\author{Mohamad Dolatshah \and Ali Hadian \and Behrouz Minaei-Bidgoli}

\institute{
Iran University of Science and Technology\\
\email{\{dolatshah.mohammad,ali.hadian\}@gmail.com, b\_minaei@iust.ac.ir}
}

\maketitle

\begin{abstract}
Emerging location-based systems and data analysis frameworks requires efficient management of spatial data for approximate and exact search. Exact similarity search can be done using space partitioning data structures, such as KD-tree, R*-tree, and ball-tree. In this paper, we focus on ball-tree, an efficient search tree that is specific for spatial queries which use euclidean distance. Each node of a ball-tree defines a ball, i.e. a hypersphere that contains a subset of the points to be searched. 

In this paper, we propose ball*-tree, an improved ball-tree that is more efficient for spatial queries. Ball*-tree enjoys a modified space partitioning algorithm that considers the distribution of the data points in order to find an efficient splitting hyperplane. Also, we propose a new algorithm for KNN queries with restricted range using ball*-tree, which performs better than both KNN and range search for such queries. Results show that ball*-tree performs 39\%-57\% faster than the original ball-tree algorithm.
\end{abstract}


\section{Introduction}

Nearest neighbor (NN) search search is of great importance in many data analytics applications such as Geographical Information Systems (GIS), machine learning, computer vision, and robotics. Given a query point $q$, a common task is to search for the $k$ closest points to $q$ among all points in a dataset. Similarly, one might want to get all points whose distances to $q$ are less than the radius $r$ (i.e. range queries). Such queries can be answered using a space-partitioning data structure, such as KD-tree~\cite{bentley1975multidimensional}. A KD-tree is a generalization of binary trees, in which each internal node represents a rectangular partition in the feature space and its subtree contains all data points that fall in the rectangle.

In the last two decades, various algorithms are proposed for efficient and exact nearest neighbors search, such as metric-tree~\cite{ciaccia1997m}, ball-tree~\cite{omohundro1989five,liu2006new}, cover-tree~\cite{bertin2011million}, VP-tree~\cite{shinde2010similarity}, MVP-tree~\cite{borodin2003lower},  R-tree~\cite{guttman1984r}, and R*-tree~\cite{beckmann1990r}. These space-partitioning data structures can be applied to either the original data or on a transformed version of the dataset.
Exact nearest neighbor search  algorithms leverage different space-partitioning methods that split the feature space using hyper-triangular or hyper-spherical bounding boxes, and build up a search tree on the resulting hierarchy of partitions. 

Space-partitioning data structures build a search tree for the given data points. The root partition contains the whole search space, and each branch of the tree splits its partition into two sub-partitions. Various space-partitioning data structures are mainly different in three aspects: 1)The geometric shape of the partition, 2)The space-partitioning strategy that builds up the search tree, and 3)The search algorithm which exploits the data structure to search for the given query.

The geometrical shape of the partitions is a key property of space-partitioning data structures that affects their performance. For example, while KD-trees and R-trees are widely used for finding nearest neighbors, they are not efficient for many data distributions due to their hyper-rectangular partition shapes, specially when the data points fall in the \textit{corners} of the hyper-rectangles~\cite{witten2005data}. As a result, the efficient solution is to use metric-trees, in which the space partitioning is specifically optimized for the distance function being used. 

In this paper, we focus on building efficient metric-trees for euclidean distances, also known as ball-trees. In a ball-tree, each internal node represents a spherical partition which allows more flexibility for space partitioning. While ball-trees have been successfully applied to various trajectory search problems, specially in computer vision, few works have been done on optimizing the space partitioning algorithms. In the ball-tree algorithm introduced by Andrew Moore~\cite{moore2000anchors}, each node is split using two steps: 1)choosing farthest point from centroid of points as the centroid of the first sub-partition, and 2)selecting the farthest point from the first one as the centroid of the second sub-partition. While this is a simple and effective procedure, it can lead to unbalanced trees. We suggest a new node splitting method, called ball*-tree, that makes more balanced trees and thus accelerates nearest neighbor search. 

In this work, we propose an efficient ball-tree that can speed up its current implementations. We propose modifications on ball-tree's space partitioning method, as well as its search algorithm. Our contributions are as follows: 1)We suggest a novel method for selecting the splitting axes in each node using principal component analysis 2)We propose heuristics for finding the best splitting hyperplane with respect to the selected axes. 3)In the search stage, we suggest an efficient algorithm for range-constrained NN search. 

The remainder of this paper is organized as follows. Section~\ref{sec:related_work} describes works which have already been done in the spatial indexes literature and introduces the metric-tree data structures. Section~\ref{sec:space_partitioning} describes methods which various types of ball-tree use for splitting space and their advantages and disadvantages then introduces our method for dividing space. Section~\ref{sec:search} describes the different kinds of algorithms that are suitable for search by metric-tree methods and their corresponding drawbacks and our solution for search the tree in a more efficient way. In Section~\ref{sec:experimental_results} we compare our entire solution with already existed ones and demonstrate the results of experimental tests on various datasets.

\section{Preliminaries and related work}
\label{sec:related_work}
Spatial indexes are very important for indexing and retrieval of trajectory data. Due to extensive development of applications in GIS and computer vision, many DBMS engines support spatial range search on geographic data. Spatial search queries are usually in form of \textbf{range} or \textbf{nearest-neighbor} search. While it is conceptually simple, efficient and effective nearest neighbor search is a very hard problem and is extensively studied in the data management communities~\cite{chavez2001searching,zezula2006similarity,naidan2015permutation}. Exact range search requires efficient algorithms for partitioning the feature space to build a balanced search tree, along with a sound search algorithm for finding all data points that reside in the range of the given query. There are two approaches for exact search: 1)\textit{Compact partition indexes} that directly partition the data points, such as KD-tree, cover-tree, and ball-tree, and 2)\textit{Pivot-based indexes} that use a set of points, called \textit{pivots}, to map the points to another space in which the distance is easier to compute~\cite{naidan2015permutation,amato2014some}. While pivot-based approaches are faster in medium-sized datasets, the required number of pivots is extremely large for high-dimensional datasets~\cite{tellez2013succinct}. 

Exact search methods leverage the triangular inequality to ensure that no relevant result is missed. There is no de-facto algorithm that can process all queries with the best performance. For example, search trees with hyper-rectangular partitions, such as R*-tree, are very efficient for queries with single-dimensional constraints, i.e. \textit{SELECT $X$ WHERE $((-3\leq x_1 \leq 1)$ AND $(11\leq x_2 \leq 25))$}. On the other side, ball-tree is efficient for nearest-neighbor search with euclidean distances.

The splitting operation has a crucial role in the efficiency of the search tree. In general, there are two approaches for splitting a partition (space) into two sub-partitions, namely \textit{binary} and \textit{multi-dimensional} splitting. In a binary split, only a single dimension is considered in each split. In a 2-dimensional space, for example, the binary splitting hyperplane is parallel to either the X or Y dimension. Most of the space-partitioning data structures are binary, such as KD-tree~\cite{friedman1977kd}, R-tree~\cite{guttman1984r}, oct-trees~\cite{samet1984oct}, quad-trees~\cite{finkel1974quad}, etc. However, in multidimensional methods, such as metric-tree and ball-tree, the split criteria is more flexible and allows considering values of multiple or all dimensions in the splitting criteria. Also, the constraints on the splitting criteria, e.g. being binary or multi-dimensional, define the shape of the resulting partitions. For example, binary splits result in rectangular partitions, while ball-trees have spherical partitions. A great disadvantage of binary split is that skewed datasets result in appearing long skinny rectangles which can lead to more number of backtrack levels during search and also producing highly unbalanced trees. Furthermore rectangles, even squares, are not the best shape for partitioning, because if a target point falls into a corner of a rectangle we would have to trace over a great number of nodes around the corner in order to find nearest neighbor which obviously cause an increment in complexity of search algorithm. Since the best split over the data is usually not binary, the search trees of multi-dimensional splits result in better performance~\cite{kumar2008what, bhatia2010survey,kibriya2007empirical}.

Several types of metric trees has been recently proposed. We first introduce the idea presented by Omohundro \cite{omohundro1989five} and Uhlmann \cite{uhlmann1991metric}. Unlike KD-trees, metric-trees do not require data to be in vector form. Hence, metric-trees can be applied to any data representation as long as the data is in the metric space~\cite{mico1994new}. For a detailed performance evaluation against established NN search methods, see~\cite{munaga2012performance,bhatia2010survey,kibriya2007empirical}.

Ball-trees are special but well-known metric-trees that use euclidean norm as the distance function. Ball-trees have been of interest to researchers for high-dimensional exact search~\cite{liu2003efficient,liu2004investigation,liu2006new}. 

A ball-tree is a binary tree in which every node defines a D-dimensional hypersphere, or ball, containing a subset of the points to be searched. Each node of the tree represents a ball, that is a hyper-spherical partition (e.g. a circle in 2D space). While the balls themselves may intersect, each point is assigned to one or the other ball in the partition according to its distance from the ball's center. 

\section{Space partitioning}
\label{sec:space_partitioning}
In this section, we introduce the ball-tree space partitioning algorithm. Then, we describe our modifications on ball-tree then optimize the efficiency of the search tree. Since there are multiple implementations for ball-tree, which are slightly different in detail, we focus on the implementation suggested by Moore~\cite{moore2000anchors}. For the sake of simplicity, we refer to Moore's ball-tree as ball-tree.

In a ball-tree, a partition is defined as the smallest hypersphere (i.e. smallest circle in 2-dimensional space) that contains all the data points belonging to that partition. We can represent a partition $p_i$ by the center $c_i$ and diameter $d_i$ of its hypersphere. A data point $\mathbf{x}$ belongs to a partition $p_i$ if $||\mathbf{x}-c_i|| < d/2$, where $||\cdot||$ is the euclidean norm. 

The ball-tree is a binary tree in which each node of the tree represents a partition. For each two nodes $v_i, v_j$, if $v_i$ is a parent of $v_j$, then the partition $p_j$ is a sub-partition of $p_i$. 

\subsection{Ball-tree space partitioning}
Ball-tree is built using a divide-and-conquer approach. Initially, ball-tree has only one (root) node and all data points are assigned to it. In each step, the partition corresponding to each node is split into two sub-partitions. For a partition $p_i$, the splitting procedure is as follows:

\begin{enumerate}
\item Select the farthest point from centroid of node points in $p_i$ as the first (left) child pivot $p^L_i$.
\item Select the farthest point from $p^L_i$ as the second (right) child pivot $p^R_i$.
\item Assign each of the data points in $p_i$ to the partition whose pivot is closer.
\item Assign the new sub-partitions as children of $v_i$ in ball-tree, i.e. $v^L_i$ and $v^R_i$.
\end{enumerate}

\begin{figure}
\centering
\includegraphics[width=0.4\textwidth]{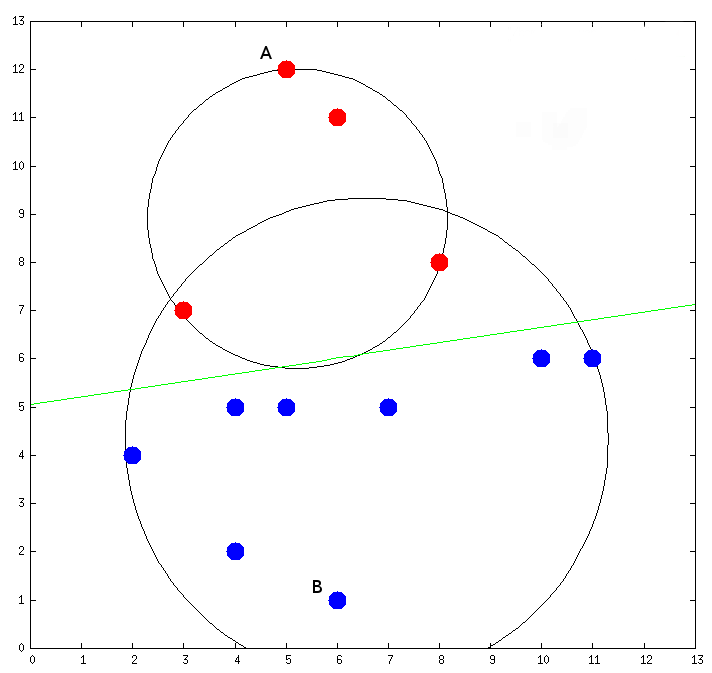}
\caption{Partitioning of a 2D point set in ball-tree. The green (dotted) line is the separating hyperplane. }
\label{fig:ball-tree-partitioning}
\end{figure}

The computational complexity of each split is $O(n)$ where $n$ indicates the number of data points in the parent partition. Figure~\ref{fig:ball-tree-partitioning} illustrates how the data points are split into two partitions according to the farthest data points ($A$,$B$). The green dotted line shows the splitting hyperplane. The points that are above the hyperplane are closer to point $A$ and thus belong to the first partition, and vice versa.

\subsection{Ball*-tree}
Ball-tree splits have two shortcomings: First, when splitting a partition, the number of points assigned to each sub-partition is not taken into account. As a result, the partitions do not necessarily have the same number of data points and the resulting tree is unbalanced. Second, the distribution of the points is not considered for defining the separating hyperplane, but the hyperplane is determined by only the two farthest points. This makes ball-tree very sensitive to outlier data points.

\begin{figure*}
    \centering
    \begin{subfigure}[b]{0.45\textwidth}
        \centering
        \includegraphics[width=\textwidth]{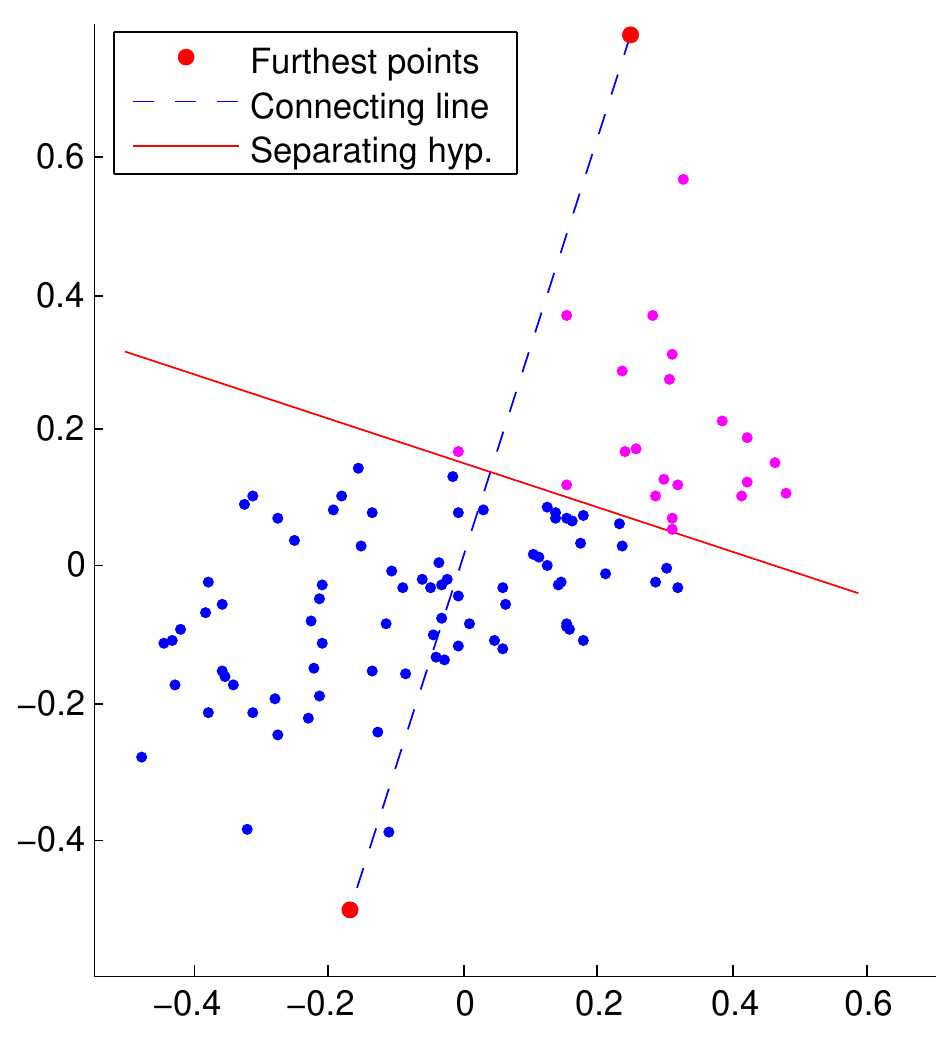}
        \caption{Ball-tree}
        \label{fig:comparison_bt_pca:bt}
    \end{subfigure}
    \hfill
    \begin{subfigure}[b]{0.45\textwidth}
        \centering
        \includegraphics[width=\textwidth]{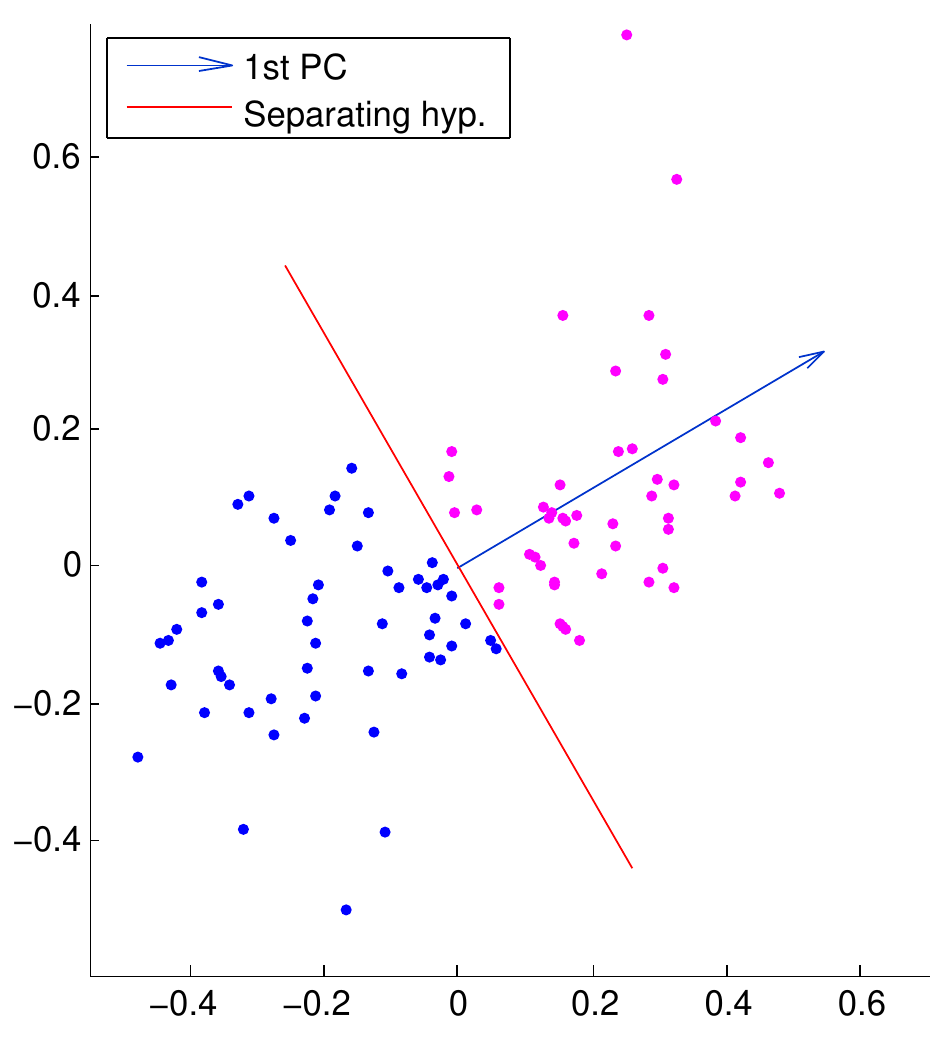}
        \caption{Ball*-tree}
        \label{fig:comparison_bt_pca:pca}
    \end{subfigure}
    \caption{Comparison of the splitting algorithms in ball-tree and ball*-tree}
    \label{fig:comparison_bt_pca}
\end{figure*}

In Ball*-tree, we address these shortcomings by taking into account the data distribution when determining the splitting hyperplane. The intuition behind ball*-tree is to detect the best direction for splitting the data points. This direction is simply extracted by the first principal component. Figure~\ref{fig:comparison_bt_pca} illustrates the difference between the splitting algorithm in ball-tree and ball*-tree. In ball-tree (Figure~\ref{fig:comparison_bt_pca:bt}), the splitting hyperplane (red line) is determined by the line that connects the two furthest points (dashed line). Therefore, the split might create unbalanced sub-partitions. In ball*-tree, as shown in Figure~\ref{fig:comparison_bt_pca:bt}, the splitting hyperplane is perpendicular to the first principal component. By detecting the direction of the data, ball*-tree's splits are more balanced, hence the resulting tree is more efficient.

In ball*-tree, the hyperplane is determined in three steps:
\begin{itemize}

\item Apply Principal Component Analysis (PCA) and find the most significant (first) eigenvector $\mathbf{w}_{(1)}$.
\item Map the data points to the axis corresponding to $\mathbf{w}_{(1)}$.
\item Find a hyperplane that is perpendicular to $\mathbf{w}_{(1)}$ and splits the data points in a balanced manner. The splitting criteria is discussed further.
\end{itemize}

\subsubsection{Data transformation}
Principal component analysis (PCA) is a well-known method for eigenvector analysis which is very effective for dimensionality reduction~\cite{abdi2010principal}. In PCA, data points $\mathbf{X} = \lbrace \mathbf{x_1}, \ldots, \mathbf{x_n} \rbrace \in \mathbb{R}^d$ are linearly transformed to a new space in $\mathbb{R}^{d'}$ ($k'\leq k$), such that the $k'$ new variables are linearly uncorrelated. The extracted uncorrelated dimensions are called principal components, which are extracted from the eigenvectors of the covariance matrix of the variables in the original space. The new points $\mathbf{T} = \lbrace \mathbf{t_1}, \ldots, \mathbf{t_n} \rbrace$ are computed by projecting the data points to the principal components $\lbrace \mathbf{w}_{(1)}, \ldots, \mathbf{w}_{(k')} \rbrace$, i.e. the $j^{th}$ dimension of the transformed data point $\mathbf{t_i}$ is determined by $t_i^{(j)} = \mathbf{x_i} \cdot \mathbf{w}_{(j)}$. An interesting property of PCA is that the first few principal components with largest eigenvalues preserve almost all of the important information in the data, in terms of the total variance. 

In ball*-tree, our target is to detect the most significant direction of the data. Hence, we apply PCA to transform the data points to a single dimension. The transformation can be done using the most significant eigenvector of the data points, i.e. the eigenvector with larges eigenvalue. The most significant primary component can be computed by solving the following optimization problem~\cite{van2009dimensionality}:

$$ \mathbf{w}_{(1)} = {\operatorname{\arg\,max}}\, \left\{ \frac{\mathbf{w}^T\mathbf{X}^T \mathbf{X w}}{\mathbf{w}^T \mathbf{w}} \right\} $$

\begin{figure}
    \centering
    \includegraphics[width=.4\textwidth]{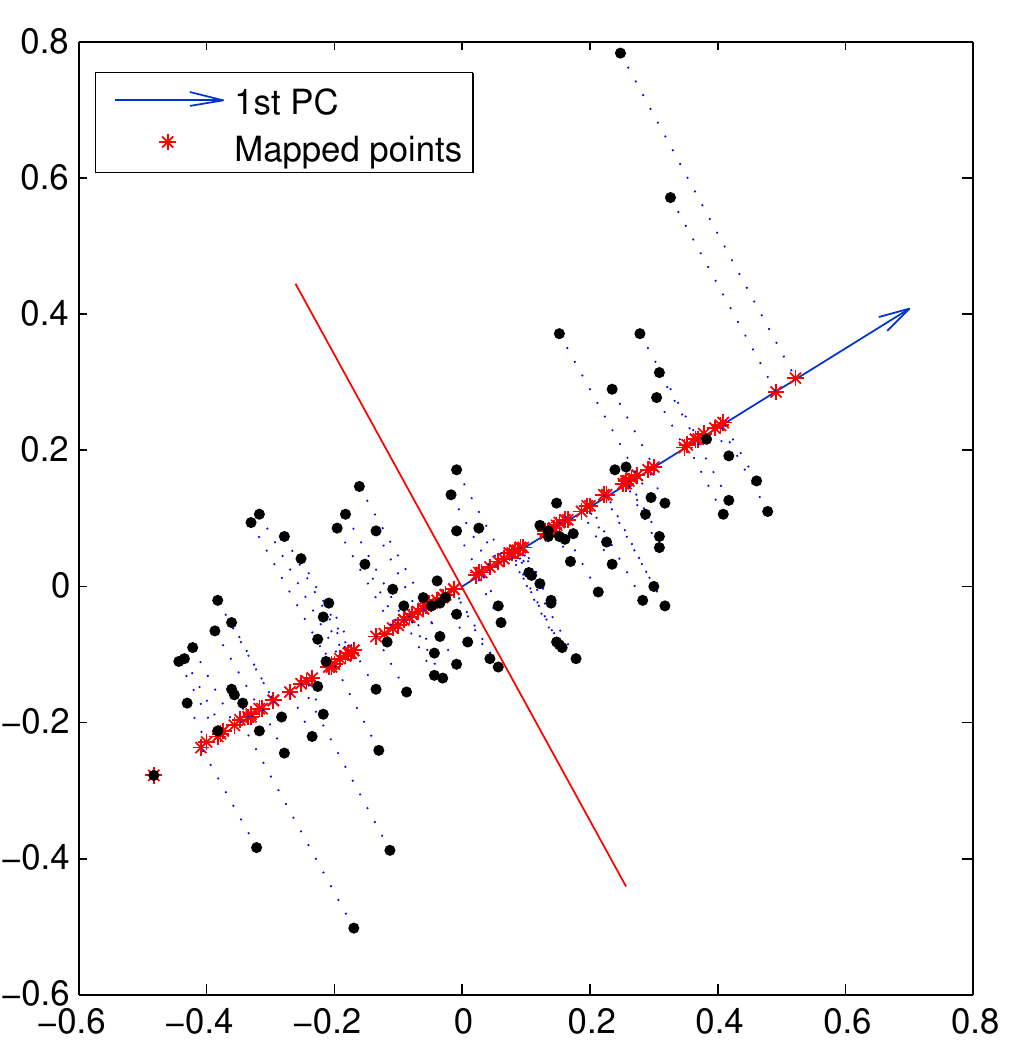}
    \caption{Mapping points to the first principal component}
    \label{fig:mapped_points}
\end{figure}

For two-dimensional data, for example, the geometric representation for $\mathbf{w}_{(1)}$ is equivalent to fitting a line to the data points using least-square minimization. Figure~\ref{fig:mapped_points} shows how the data points in a two-dimensional space are transformed to the 1-dimensional space. The cross-points indicate the projection of the data points on the most significant primary component. The dotted line show the perpendicular offsets between the ${w}_{(1)}$ and the data points. Each data point is transformed to the single-dimensional space corresponding to $\mathbf{w}_{(1)}$, as $t_i = \mathbf{x_i} \cdot \mathbf{w}_{(1)}$. 

\subsubsection{Best splitting hyperplane}
In ball*-tree, the splitting hyperplane is perpendicular to the eigenvector $\mathbf{w}_{(1)}$ and is determined by ${\mathbf{w}_{(1)} \cdot \mathbf{x} -b = 0}$. The parameter $\tfrac{b}{\|\mathbf{w}_{(1)}\|}$ determines the offset of the hyperplane from the origin along the normal vector $\frac{\mathbf{w}_{(1)}}{\|\mathbf{w}_{(1)}\|}$. An important issue is to find $b$ such that the splitting hyperplane yields two balanced partitions and the optimality of the resulting ball*-tree is maximized. This is equivalent to determining the optimal point where the splitting hyperplane intersects the axis line $\mathbf{w}_{(1)} \cdot \mathbf{x}=0$.

The splitting hyperplane should be selected in a way that reduces the search cost in the resulting tree. To define an objective function, assume that search queries have a random distribution and are independent of dataset model. In the search tree, the children of each nodes should be as small as possible, in order to go through less number of nodes for a search query. Also, the two children of a node should be as even as possible, to achieve a balanced tree. Therefore, we should simultaneously optimize two objectives:
\begin{enumerate}
\item Maximize the balance between the number of points in the child partitions ($N_1$ and $N_2$).
\item Minimize the radius of each child partition ($R_1$ and $R_2$) 
\end{enumerate}

Since $N_1+N_2=N$, the balance between two partitions is maximized when $N_1=N_2=\frac{N}{2}$. Therefore, the first objective can be formulated as minimizing $f_1=\frac{|N_2-N_1|}{N}$. More specifically, in case that the query points come from the same distribution as the original data points, the tree is optimized for such a workload if $f_1$ is maximized. 

The second objective is minimized when the two sub-partitions have equal diameters. For the sake of simplicity and efficiency, instead of finding the two furthest nodes in each sub-partition, we use the distances of the two furthest points in $T$ which are scalar values. Assume that the maximum and minimum values of $T$ are $t_{max},t_{min}$, respectively. The objective is minimized when the intersection point $t_c$ is in the middle of $t_{max}$ and $t_{min}$, i.e. ${t_c = (t_{max}+t_{min})/2}$. Therefore, the objective can be defined as ${f_2 = \frac{t_c - t_{min}}{t_{max}-t_{min}}}$. The second objective is very important if the query points come from a distribution that is different from the original data distribution. Thus, it is important to have a ball*-tree that is optimized for similar workloads as in the original data distribution (optimized by $f_1$), as well as other datasets (optimized by $f_2$).

To put it all together the best splitting plane is determined by finding $t_c$ for which the following objective function is minimized

$$F(t_c) = \frac{|N_2-N_1|}{N} + \alpha \left( \frac{t_c - t_{min}}{t_{max}-t_{min}} \right) $$.

In $F(t_c)$, $\alpha>0$ is a workload-awareness parameter, i.e. if the workloads is assumed to be from the same distribution as the original dataset, we set small $\alpha$ values to pay more attention to $f_1$.

In the above optimization problem $N_1,N_2$ should be explicitly computed, hence there is no closed-form solution that minimizes $F(t_c)$. Instead, we evaluate $F(t_c)$ for a finite set of values in ${\left[ t_{min},t_{max} \right]}$ and choose the value for which $F(t_c)$ is minimized. The range ${\left[ t_{min},t_{max} \right]}$ is split into $S$ sections to extract $S$ values, and $F(t_c)$ is evaluated for the mean value of each section.

Algorithm~\ref{alg:split} illustrates the splitting method for ball*-tree. Similar to ball-tree, the computational complexity of our suggested splitting method is $O(n)$.

\begin{algorithm}
\caption{Ball*-tree: Space partitioning}
\label{alg:split}
\begin{algorithmic}[1]
\Procedure{split}{$\mathbf{X}$}

\State \textbf{Input:} $\mathbf{X}$ (Data points in the partition)
\State \textbf{Output:} $\mathbf{X}^\mathbf{R}$, $\mathbf{X}^\mathbf{L}$ (Data points of the right/left partition)

\State $\mathbf{w}_{(1)} = {\operatorname{\arg\,max}}\, \left\{ \frac{\mathbf{w}^T\mathbf{X}^T \mathbf{X w}}{\mathbf{w}^T \mathbf{w}} \right\} $
                \Comment{Apply PCA}
\State $T = \mathbf{w}_{(1)} \cdot X$   \Comment{Transformation}

\State $t_c = {\operatorname{\arg\,min}}\, \frac{|N_2-N_1|}{N} + \alpha \left( \frac{t_c - t_{min}}{t_{max}-t_{min}} \right)$

\State $\mathbf{X}^\mathbf{R} = \lbrace \mathbf{x}_i | \mathbf{x}_i \in \mathbf{X}, t_i < t_c \rbrace$
\State $\mathbf{X}^\mathbf{L} = \lbrace \mathbf{x}_i | \mathbf{x}_i \in \mathbf{X}, t_i \geq t_c \rbrace$

\EndProcedure
\end{algorithmic}
\end{algorithm}

\section{Query processing}
\label{sec:search}
A ball-tree can be used for both \textit{range search} and and \textit{KNN search}. In range search, the result set should contain all points in the specified range, while in KNN search only the K closest points are returned. In ball*-tree search algorithm we make use of both of these methods in order to build a new algorithm for the constrained NN area.

\subsection{Range search in ball-tree}
Given a range query $(\mathbf{q},r)$, the result set contains all data points where ${||q-\mathbf{x}||\leq r}$. The search procedure starts from the root node and recursively evaluates every child node that might contain a result in the range. More specifically, for each sub-partition $p_i$ corresponding to a node in the ball-tree, if $(\operatorname{radius}(p_i)+r) \leq ||\operatorname{center}(p_i)-q||$, then the elements inside the partition are evaluated to find possible data points that satisfy $||\textbf{x}-q|| \leq r $. In other words, if the query ball intersects with the node's ball. The search is continued further to all child nodes that intersect with the query ball.\cite{leibe2006efficient}.

\begin{figure}
\centering
\includegraphics[width=.6\textwidth]{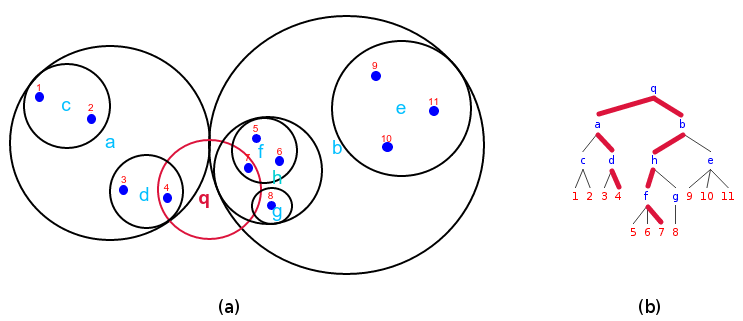}
\caption{(a) Ball-tree partitions (b) Corresponding search tree}
\label{fig:ball_tree_representation}
\end{figure}

Figure~\ref{fig:ball_tree_representation} illustrates how a range query is processed in a ball-tree. Both partitions $a,b$ intersect with $q$'s range. In the next levels, sub-partitions $d,h,f$ intersect with $q$, and the data points in leaf nodes (i.e. $d,f$) are linearly searched for possible results. Partitions $c,e,g$ are too far from the query, thus skipped. Finally, $\lbrace \textbf{x}_4,\textbf{x}_7 \rbrace$ is returned as the result set.

\subsection{K-NN search}

K-NN search always returns the K nearest neighbors of the target point. We briefly explain the KNN search method on Ball-tree, as proposed by Liu et al~\cite{liu2006new}. The algorithm considers a list of points $P^{in}$ which contains the points found so far as the nearest neighbors of the target point ($t$). Also, let $D_{s}$ be the minimum distance from target point and previously discovered nodes $ D_{s} =  \max_{x \in P^{in}} |x-t| $ , and $D_{N}$ be the distance between $t$ and current node.

$$ D_{N} =  max\{D_{N.Parent},|t-\operatorname{center}(N)|-\operatorname{radius}(N)\}  $$

In K-NN search algorithm, a node is expanded if $D_{N} < D{s}$. If the current node is a leaf, then every data point $x$ in $N$ that satisfies 
$||x-t||<D_{s}$, $x$ is added to the results list. When the KNN list size exceeds the $K$ limit, the furthest point is removed from the list and $D_{s}$ is updated for further execution.

\subsection {Constrained NN in ball*-tree}
Our key idea for implementing range-constrained K-NN search is to combine ball-tree's K-NN and range search algorithms and enjoy pruning from both. The range constraint limits the number of candidate nodes, K-NN prunes the search nodes by the nearest points found so far. In other words, whenever a node is either too far from the query (w.r.t range) or is not likely to be among the top K points found so far, it is skipped. Algorithm~\ref{alg:mixtureSearch} represents the constrained NN search algorithm for ball*-tree.

\begin{algorithm}
\caption{Constrained NN algorithm for ball*-tree}
\label{alg:mixtureSearch}
\begin{algorithmic}[1]
\Procedure{find($P^{in},node,r,K$)}{}

\If  {($D_{N} >= D_{s} \wedge D_{N} > r$)} 
\State \Return $P^{in}$ unchanged.
\EndIf

\If{\textit{node} is a leaf}
\State $P^{out}=P^{in}$
\State $\forall x\in \operatorname{points}(node)$:  \textbf{if} $(|x-t|< D_{s})$ \textbf{then}  add x to $P^{out}$

\If{$(|P^{out}|==K+1)$}
\State Remove farthest neighbor from $P^{out}$
\State Update $D_{s}$
\EndIf 
\Else
\State $d_R$ = distance from $child_R$ center
\State $d_R$ = distance from $child_L$ center
\State $P^{temp}=P^{in}$
\If{$(dR<=\operatorname{radius}(\operatorname{child}^R(node))+r)$}
\State $P^{temp}$ = FIND($P^{in}$ , $child_R$ , $r$)
\EndIf 
\If{$(dL <= \operatorname{radius}(\operatorname{child}^L(node))+r)$}
\State $P^{out}$ = FIND($P^{temp}$ , $child_L$ , $r$)
\EndIf 
\EndIf 
\EndProcedure
\end{algorithmic}
\end{algorithm}

\section{Experimental results}
\label{sec:experimental_results}

We evaluated the performance of ball*-tree on both synthetic and real-world datasets. The synthetic graphs are generated to simulate datasets with customized densities. We prepared 5 synthetic datasets with $500,000$ data points in 2-dimensional space, namely \textit{Latin-center}~\cite{colbourn2006handbook}, \textit{Highleyman}~\cite{duin2004PRTools}, \textit{Niederreiter}~\cite{niederreiter1988low}, \textit{Lithunian}~\cite{duin2004PRTools}, and \textit{Sobol}~\cite{sobol1967distribution}. Also, we compared the algorithms against two real-world datasets with $10,000$ data points in 4-dimensional space, namely \textit{Skin Segmentation Data Set}~\cite{Lichman2013} and \textit{3D Road Network Data Set}~\cite{kaul2013building}. 

Since our contribution is two fold, i.e. PCA-based partitioning and constrained NN search, we first evaluate our construction algorithm and then provide further experiments on how our constrained NN search algorithm accelerates ball*-tree for range-constrained NN queries.

\subsection{Space-partitioning}
The execution time depends on how many leaf nodes are linearly searched for possible matches that fall in the range of the given query. Therefore, we first investigate the efficiency of the comparing algorithms in terms of average depth which is the the average distance from the root node to each leaf of the tree. The average depth of a tree is desired to be minimized, which results in less number of comparisons.

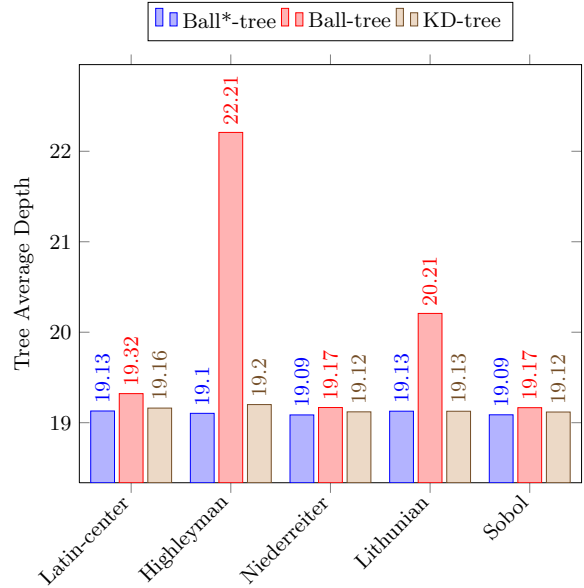
\begin{figure}
    \centering
    \begin{tikzpicture}[thick,scale=0.9]
    \tikzset{style={font=\small}}
    	\begin{axis}[ 
    			width=9cm,
				ybar, 
				enlargelimits=0.13, 
				legend style={
					at={(0.5,1.15)}, 
					anchor=north,
					legend columns=-1}, 
				ylabel={Tree Average Depth}, 
				ylabel style={yshift=-3ex},
				symbolic x coords={Latin-center,Highleyman,Niederreiter,Lithunian,Sobol}, 
				 x tick label style={rotate=45,anchor=east},
				xtick=data,
				enlarge y limits=0.24,
				ytick={19,20,21,22,23},
				nodes near coords, 
				every node near coord/.append style={rotate=90,anchor=west},
				bar width=10pt,
				] 
    		 
    		\addplot coordinates {(Latin-center,19.1294) (Highleyman,19.1035) (Niederreiter,19.0859) (Lithunian,19.1272) (Sobol,19.0873)}; 
			\addplot coordinates {(Latin-center,19.3214)(Highleyman,22.2084) (Niederreiter,19.1677) (Lithunian,20.2078) (Sobol,19.166)};
			\addplot coordinates {(Latin-center,19.1615)(Highleyman,19.2003) (Niederreiter,19.12) (Lithunian,19.1266) (Sobol,19.118)};
    		
    		\legend{Ball*-tree,Ball-tree,KD-tree} 
    	\end{axis} 
	\end{tikzpicture}
    \caption{Comparison of the average depth}
    \label{fig:comparison_bt_b*t}
\end{figure}

Figure~\ref{fig:comparison_bt_b*t} compares the average path length of ball*-tree, ball-tree, and KD-tree. Results show that average depth in ball*-tree is shorter than ball-tree, but almost equal to KD-tree. 

In order to compare the efficiency of our space partitioning method, we use the same search algorithm for ball-tree and ball*-tree.

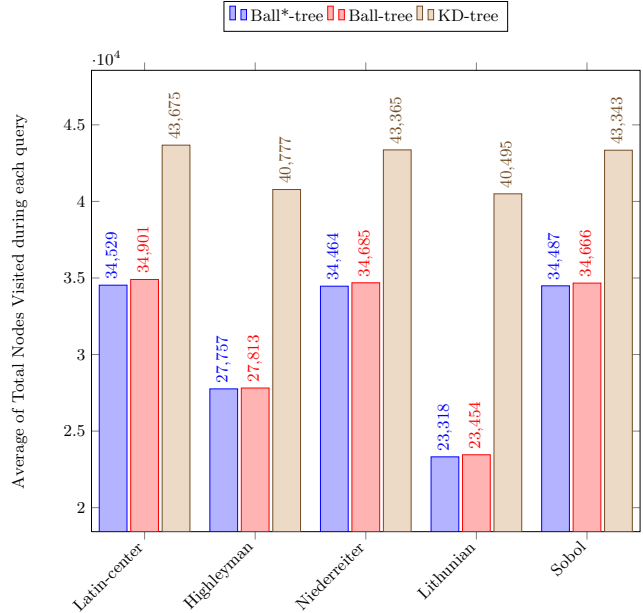
\begin{figure}
\centering
\begin{tikzpicture}[thick,scale=0.7]
    \tikzset{ style={}}
    	\begin{axis}[
    			width=12cm, 
				ybar, 
				enlargelimits=0.12, 
				legend style={
					at={(0.5,1.15)}, 
					anchor=north,
					legend columns=-1}, 
				ylabel={Average of Total Nodes Visited during each query},
				ylabel style={yshift=1ex}, 
				symbolic x coords={Latin-center,Highleyman,Niederreiter,Lithunian,Sobol}, 
				 x tick label style={rotate=45,anchor=east},
				xtick=data,
				enlarge y limits=0.24,
				nodes near coords, 
				nodes near coords, 
				every node near coord/.append style={rotate=90,anchor=west},
				bar width=15pt,
				ytick style={font=\tiny} ] 
    		\addplot coordinates {(Latin-center,34529) (Highleyman,27757) (Niederreiter,34464) (Lithunian,23318) (Sobol,34487)}; 
    		\addplot coordinates {(Latin-center,34901) (Highleyman,27813) (Niederreiter,34685) (Lithunian,23454) (Sobol,34666)}; 
    		\addplot coordinates {(Latin-center,43675) (Highleyman,40777) (Niederreiter,43365) (Lithunian,40495) (Sobol,43343)};

    		\legend{Ball*-tree,Ball-tree,KD-tree} 
    	\end{axis} 
\end{tikzpicture}
\caption{Average of number of nodes visted during each query \label{fig:visited_nodes}}
\end{figure}

The query set contains $5300$ instances with 2 numerical attributes. The query points are randomly drawn with uniform distribution in same range of values in each dataset.
Figure~\ref{fig:visited_nodes} demonstrates average number of nodes visited for each query point. Results show that both ball-tree and ball*-tree visit less nodes than KD-tree, which is a result of inherent multi-dimensional splitting criteria in ball-tree. Also, ball-tree visits between 100 to 400 nodes more than ball*-tree for each query. The same experimental is done for real-world datasets and results are shown in Table~\ref{table:real}.

\begin{table*}
    \caption{Comparison of the average depth for the real-world datasets\label{table:real}}
    \centering
     \begin{tabular}{lccc}
        \toprule
        Dataset & Ball*-tree   & Ball-tree & KD-tree \\
        \midrule
        Skin Segmentation & 13.79 & 15.76 & 13.57 \\
        3D Road Network & 14.16 & 14.78 & 13.42 \\
     \end{tabular}

\end{table*}

\subsection{Constrained NN search}

In this section, we compare our proposed constrained NN search algorithm with with the K-NN search algorithm by Liu et al~\cite{liu2006new}. For the sake of fairness, we use ball*-tree's space-partitioning alorithm for both of the competing methods. 
Table~\ref{table:visited_nodes} shows average number of nodes visited for each query point by each search algorithm.

\begin{table*}
    \caption{\label{table:visited_nodes}Average number of nodes visited for each query}
    \centering
     \begin{tabular}{llllll}
        \toprule
        & latin-center & Highleyman & Niederreiter & lithunian & Sobol \\
        \midrule
        Ball-tree    & 749.26 & 4547.35 & 662.65 & 1563.12 & 740.35\\
        Ball*-tree 	 & 102.18 & 375.63 & 100.75 & 60.84 & 81.65\\
     \end{tabular}
\end{table*}

\begin{figure*}
    \centering
    \begin{subfigure}[b]{0.45\textwidth}
	\begin{tikzpicture}[thick,scale=0.7]
		\begin{axis}[stack plots=y,
		ylabel={Search Time ( milliseconds ) }, 
		symbolic x coords={Latin-center,Highleyman,Niederreiter,Lithunian,Sobol} ,
		x tick label style={font=\tiny,rotate=45,anchor=east}]
	
		\addplot coordinates
			{(Latin-center,236721) (Highleyman,2577548) (Niederreiter,192048) (Lithunian,1192512)(Sobol,232414)};
		
		\addplot coordinates
			{(Latin-center,235617) (Highleyman,1550290) (Niederreiter,97880) (Lithunian,510058)(Sobol,117178)};

			\legend{Ball*-tree,Ball-tree} 
		\end{axis}
	\end{tikzpicture}
	\caption{\label{fig:search_time}}
	\end{subfigure}
    \hfill
    \begin{subfigure}[b]{0.45\textwidth}
    \begin{tikzpicture}[thick,scale=0.7]
		\begin{axis}[stack plots=y,
		ylabel={Search Time ( milliseconds ) }, 
		symbolic x coords={$10^3$,$10^4$,$10^5$,$10^6$,$10^7$},
		ytick = {1,2,3,4,5,6},
		yticklabels = {$10$,$10^2$,$10^3$,$10^4$,$10^5$,$10^6$},
		xlabel={Number of points } ,
		x label style={font=\large}]
		]
	
		\addplot coordinates
			{($10^3$,1.2) ($10^4$,1.66) ($10^5$,2.589) ($10^6$,4.2876)($10^7$,6.97475)};
		\node at (axis cs:$10^3$,2) [anchor=north] {12};
		\node at (axis cs:$10^4$,2.5) [anchor=north] {83};
		\node at (axis cs:$10^5$,3.5) [anchor=north] {863};
		\node at (axis cs:$10^6$,5) [anchor=north east] {10719};
		\node at (axis cs:$10^7$,7) [anchor=east] {559495};	
							
		\end{axis}
	\end{tikzpicture}
	\caption{\label{fig:scalability}}
	\end{subfigure}
\caption{(a) Ball*-tree and Ball-tree Search time (b) Scalability of Ball*-tree}
\end{figure*}
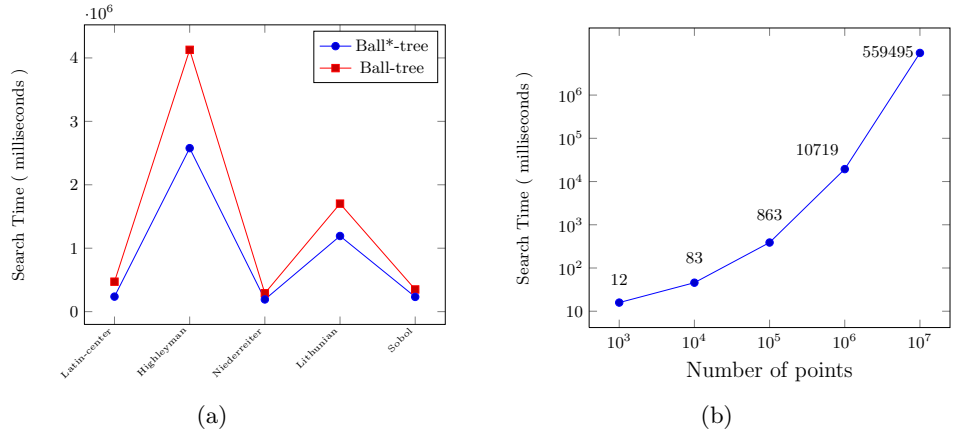

The number of visiting nodes is reduced by our method (more than 50\% in some cases).
This is illustrated in Figure~\ref{fig:search_time}. Again in some cases our algorithm managed to reduce the search time by more than 50\%.

Figure~\ref{fig:scalability} represents scalability of our algorithm in which the growth of search time based on the number of points in dataset is shown.

\section{Conclusion and Future work}
In this paper, we proposed a modified ball-tree algorithm for constrained and exact NN search. Our space-parittioning algorithm considers the distribution of the data in order to find a proper hyperplane that minimizes the effort to find the target point in the resulting tree. The hyperplane is determined by PCA, along with an optimization function that simultaneously maximizes the balance between sub-partitions as well as their average radius. Also, we proposed a new search algorithm for NN search where the results points are constrained to be in a given range from the query point. Our search algorithm is a hybrid method that combines range search and KNN search in ball-trees. Finally, we compared our proposed method with other existing spatial indexing data structures, namely ball-tree and KD-tree. 

\paragraph{Future works.}
Due to the huge amount of data to be processed, spatial indexing data structures require efficient parallel and distributed implementations for modern hardware, such as multi-core systems and General-Purpose GPUs (GPGPUs). Nevertheless, exploiting the massive parallelism offered by such hardware is challenging for spatial indexing algorithms. Efforts have been done for integrating GPGPU technology and KD-tree structure \cite{zhou2008real, zhou2014build, danilewski2010binned}, but no work have been done ball-tree construction on GPGPU, which makes it an open issue for further attention.

\bibliographystyle{splncs} 
\bibliography{ball_tree}

\end{document}